\documentclass[letterpaper]{article} 
\usepackage{aaai24}  
\usepackage{times}  
\usepackage{helvet}  
\usepackage{courier}  
\usepackage[hyphens]{url}  
\usepackage{graphicx} 
\usepackage{natbib}  
\usepackage{caption} 
\usepackage{algorithm}
\usepackage{algorithmic}

\usepackage{newfloat}
\usepackage{listings}
\DeclareCaptionStyle{ruled}{labelfont=normalfont,labelsep=colon,strut=off} 
\floatstyle{ruled}
\newfloat{listing}{tb}{lst}{}
\floatname{listing}{Listing}
\usepackage{amsmath}
\usepackage{subfigure}
\usepackage{enumitem}
\usepackage{amsfonts}
\usepackage{xspace}
\usepackage{booktabs}
\usepackage{multirow}
\newcommand{\name}{HierRec\xspace}
\newcommand*\samethanks[1][\value{footnote}]{\footnotemark[#1]}

\title{Scenario-Aware Hierarchical Dynamic Network for Multi-Scenario Recommendation}
\author{
    Jingtong Gao\equalcontrib\textsuperscript{\rm 1},
    Bo Chen\equalcontrib\textsuperscript{\rm 2},
    Menghui Zhu\equalcontrib\textsuperscript{\rm 2},
    Xiangyu Zhao\textsuperscript{\rm 1}\thanks{Correspongding Author},
    Xiaopeng Li\textsuperscript{\rm 1},\\
    Yuhao Wang\textsuperscript{\rm 1},
    Yichao Wang\textsuperscript{\rm 2},
    Huifeng Guo\textsuperscript{\rm 2},
    Ruiming Tang\textsuperscript{\rm 2}\samethanks[2]
}
\affiliations {
    \textsuperscript{\rm 1}City University of Hong Kong\\
    \textsuperscript{\rm 2}Huawei Noah’s Ark Lab\\
    \{jt.g, xiaopli2-c, yhwang25-c\}@my.cityu.edu.hk,
    xianzhao@cityu.edu.hk,\\
    \{chenbo116, zhumenghui1, wangyichao5, huifeng.guo, tangruiming\}@huawei.com\\
}

\begin{document}

\maketitle

\begin{abstract}

    Click-Through Rate (CTR) prediction is a fundamental technique in recommendation and advertising systems. Recent studies have shown that implementing multi-scenario recommendations contributes to strengthening information sharing and improving overall performance.
    However, existing multi-scenario models only consider coarse-grained explicit scenario modeling that depends on pre-defined scenario identification from manual prior rules, which is biased and sub-optimal. 
    To address these limitations, we propose a Scenario-Aware \textbf{Hier}archical Dynamic Network for Multi-Scenario \textbf{Rec}ommendations (\name), which perceives implicit patterns adaptively and conducts explicit and implicit scenario modeling jointly.
    In particular, HierRec designs a basic scenario-oriented module based on the dynamic weight to capture scenario-specific information. Then the hierarchical explicit and implicit scenario-aware modules are proposed to model hybrid-grained scenario information. 
    The multi-head implicit modeling design contributes to perceiving distinctive patterns from different perspectives.
    Our experiments on two public datasets and real-world industrial applications on a mainstream online advertising platform demonstrate that our HierRec outperforms existing models significantly.

\end{abstract}

\section{Introduction}\label{sec:1}

Click-Through Rate (CTR) prediction is a fundamental technique for online advertising and recommender systems~\cite{richardson2007predicting, yang2022click, zhang2021deep,gao2023autotransfer}. 
To improve the prediction accuracy and mitigate the data sparsity, the multi-scenario recommendation (a.k.a., multi-domain recommendation) is proposed by aggregating samples of similar scenarios~\cite{sheng2021one} for training a unified model jointly. Specifically, samples from different scenarios are explicitly distinguished by a newly introduced feature usually called ``Scenario ID'', which is manually pre-defined based on the scenario characteristics (e.g., different advertising slots or channels on the same platform). 
By modeling the connections between these scenarios, multi-scenario recommendation contributes to strengthening information sharing among different scenarios and improving the prediction effect of overall scenarios.

\begin{figure}[h]
    \vspace{-3mm}
    \centering
    \includegraphics[width=1.0\linewidth]{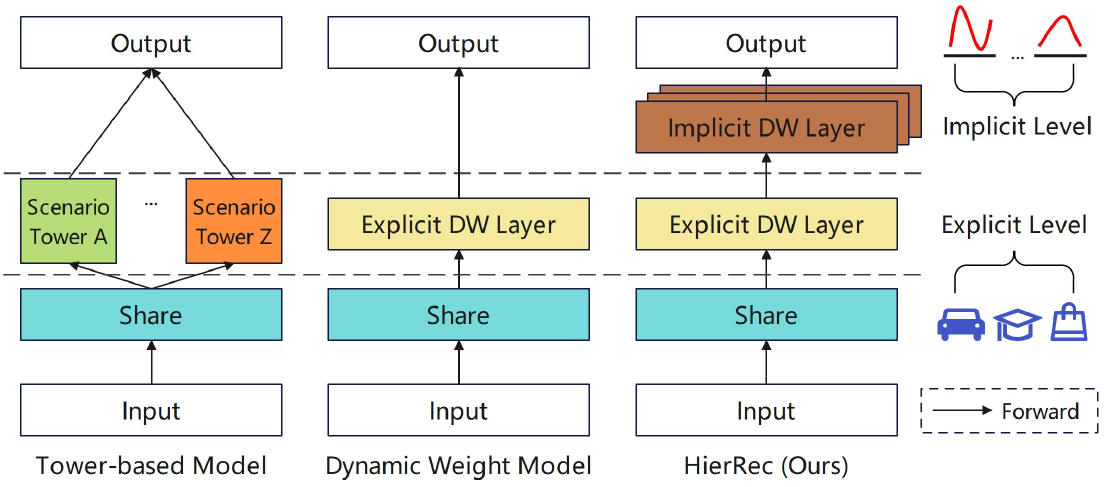}
    \caption{Comparison of different multi-scenario models.}
    \label{fig:compare}
    \vspace{-1mm}
\end{figure}

The core challenge of multi-scenario modeling is to portray scenario similarities and differences accurately. Based on the different architectures, the existing multi-scenario models can be divided into two categories: Tower-based models and Dynamic Weight (DW) models, whose abstract structures are depicted in the left part of Figure~\ref{fig:compare}.
Tower-based models leverage a shared bottom network to model scenario-shared information, based on which several sub-towers are stacked to capture scenario-specific information~\cite{sheng2021one,wang2022causalint}.
However, the design of complete isolation between towers hinders the fine-grained modeling for scenario correlations. Besides, these methods have poor generalization and compatibility when facing a large number of scenarios.
To overcome these limitations, the DW-based methods are proposed by generating dynamic parameters adaptive for each scenario in a parameter-efficient manner~\cite{zhang2022leaving, yang2022adasparse}, thus solving the generalization problem and facilitating the modeling of correlations between scenarios.

\begin{figure}[t]
	\centering
        \subfigure[The number of samples.]{
		\label{fig:dis1}
		\includegraphics[width=0.47\linewidth]{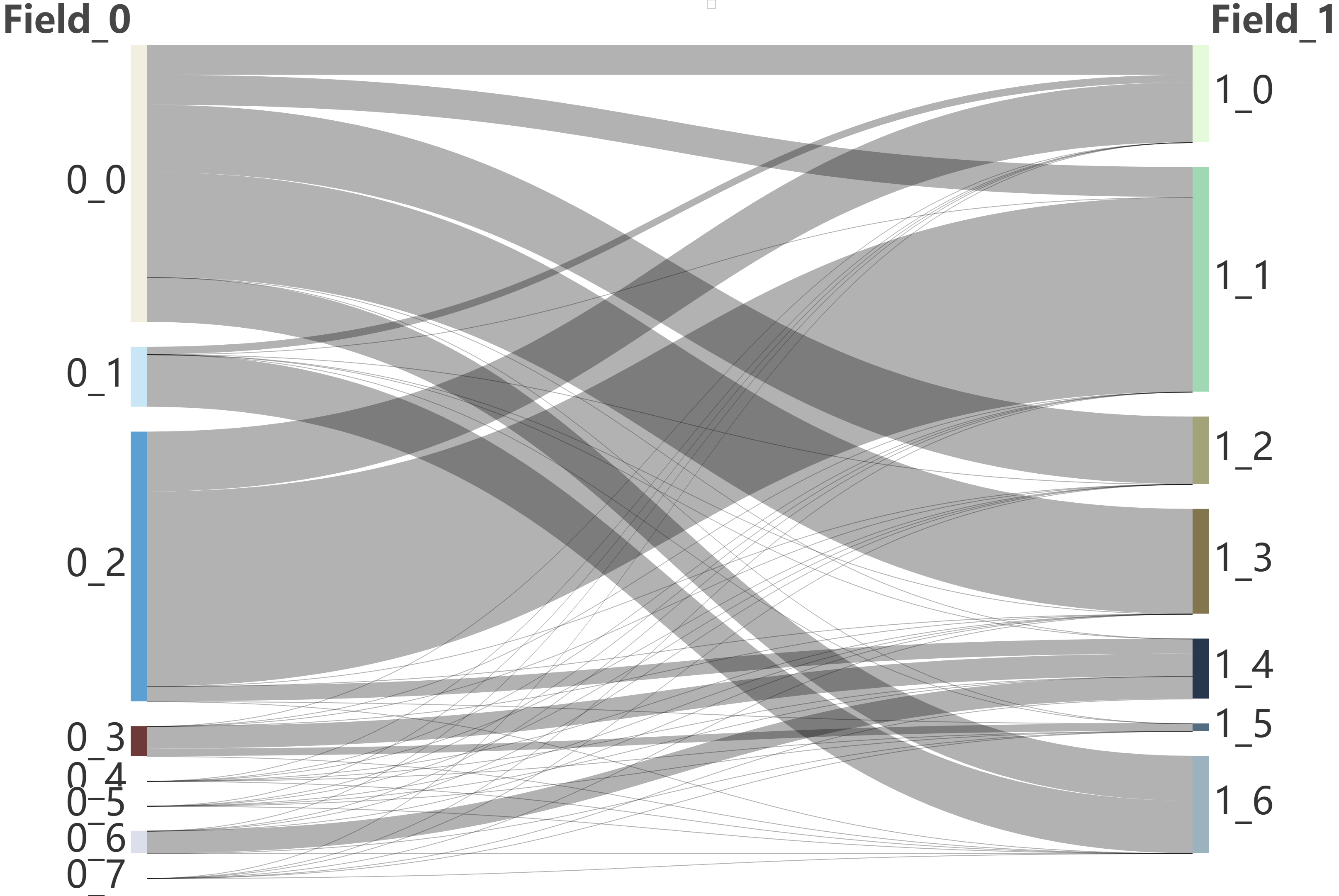}}
        \subfigure[Click-Through Rate.]{
		\label{fig:dis2}
		\includegraphics[width=0.49\linewidth]{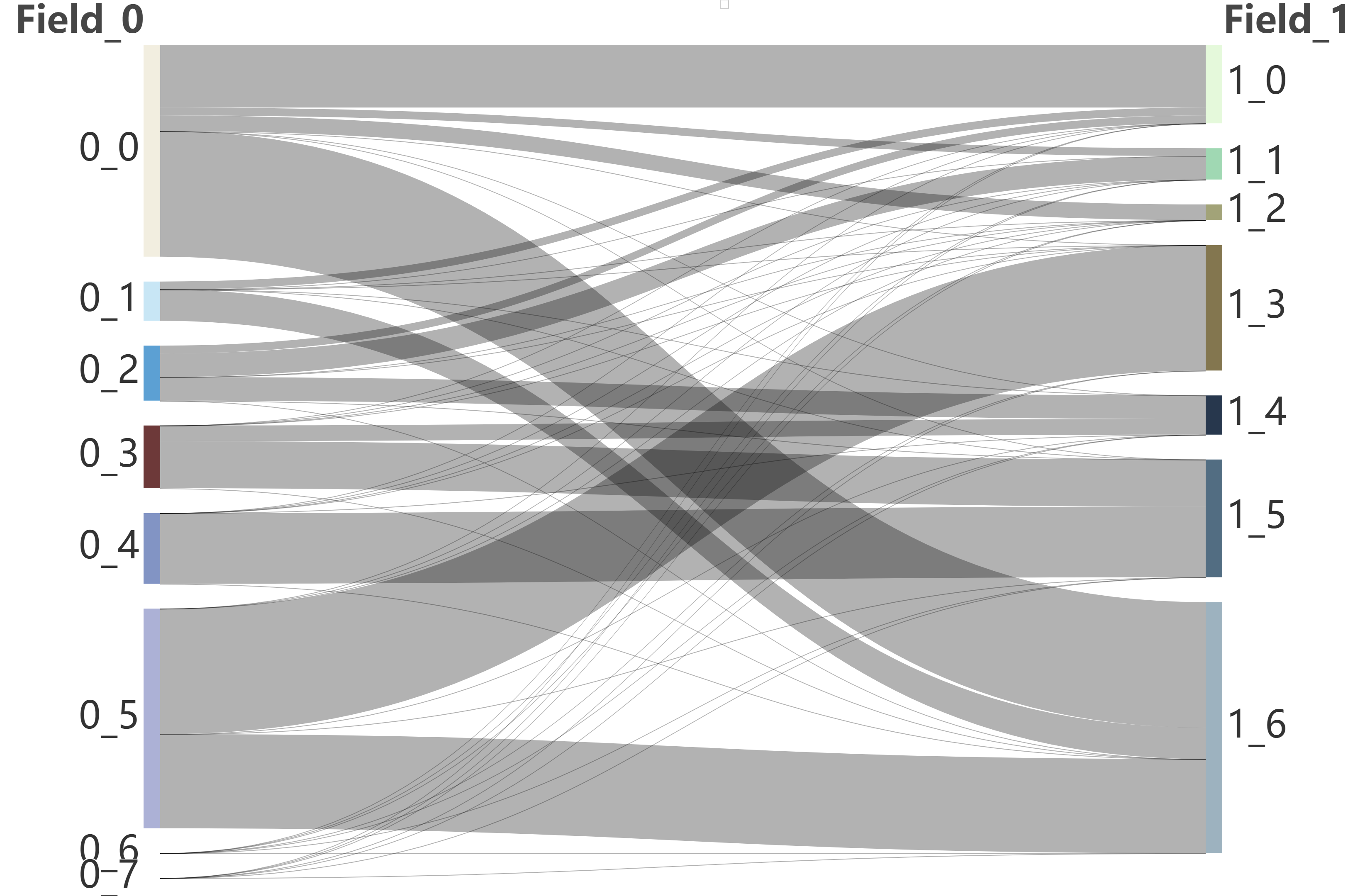}}
  \vspace{-1mm}
\caption{The number of samples and Click-Through Rate within an explicit scenario in the KuaiRand dataset. The wider flow in the Sankey diagram represents a larger number of samples or a higher Click-Through Rate.}

\vspace{-6mm}
\label{fig:distribution}
\end{figure}

However, existing multi-scenario models only consider \textbf{explicit scenario} modeling that depends on pre-defined scenario identification based on manual prior rules (e.g., advertising slots or channels) and ignore data differences within scenarios, which is biased and sub-optimal~\cite{bian2022can,wang2022causalint}. 
Taking two feature fields in the KuaiRand dataset~\cite{gao2022kuairand} as an example (shown in Figure~\ref{fig:distribution}), we can observe that the data distribution (e.g., the number of samples and the Click-Through Rate) within an explicit scenario varies markedly under different features, especially for feature combinations. 
These features or feature combinations implicitly and meticulously divide the data into various sub-scenarios where the data distribution is closer in some dimensions.
Distinguishing and utilizing these feature-based implicit patterns (referred to as \textbf{implicit scenarios}) for fine-grained modeling would thus greatly uncover more intricate correlations among different samples.
However, existing multi-scenario models~\cite{wang2022causalint} neglect the differences in these feature-based implicit patterns, hindering the recommendation performance.
Therefore, it is crucial to explore implicit scenarios and conduct detailed modeling for multi-scenario recommendations.
To achieve this, two major challenges need to be solved:
1) \textit{How to combine explicit modeling with implicit modeling in multi-scenario recommendations?}
and 2) \textit{How to perceive implicit patterns adaptively and conduct fine-grained modeling?}

To address challenges above, we propose a Scenario- Aware \textbf{Hier}archical Dynamic Network for Multi-Scenario \textbf{Rec}ommendations (HierRec), which is a hierarchical structure with an explicit scenario-oriented layer and several implicit scenario-oriented layers, shown in the right part of Figure~\ref{fig:compare}.
Specifically, HierRec first designs a scenario-oriented module based on the dynamic weight to capture scenario-specific information.
Based on this basic module, an explicit scenario-aware module is proposed to model coarse-grained explicit scenario information. Then an implicit scenario-aware module is leveraged to perceive distinctive implicit patterns and conduct fine-grained scenario modeling.
HierRec proposes a scenario-aware multi-head attention structure to identify important implicit patterns in a soft-selection manner. Subsequently, several implicit scenario-oriented layers are deployed parallelly to capture complicated distributions, thus facilitating fine-grained implicit scenario modeling.
Our contributions in this paper can be summarized as follows:
\begin{itemize}[leftmargin=*]
    \item To the best of our knowledge, this is the first work considering both explicit scenario and implicit scenario modeling in multi-scenario recommendations;
    \item We propose a multi-scenario model \name based on the dynamic weight, where stacked explicit and implicit scenario-aware modules are proposed to capture explicit and implicit information, respectively.
    Besides, multi-head implicit modeling design contributes to perceiving complicated distribution;

    \item Comprehensive experiments on two public benchmark datasets and applications on a mainstream online advertising platform demonstrate that \name outperforms existing multi-scenario recommendation models significantly.
\end{itemize}

\section{Method}

In this section, we first describe the problem formulation of the multi-scenario CTR prediction, and then provide an overview of \name and detail its key components. 

\subsection{Problem Formulation}\label{sec:problem}


Considering a training dataset $\mathcal{D}=\left\{\left(x_j, y_j\right)\right\}_{j=1}^{|\mathcal{D}|}$ with $|\mathcal{D}|$ samples, where $\boldsymbol{x}_j=\{s, c_1,...c_i,...c_I \}$ and $y_j$ represent the feature set and binary click label of the $j_{th}$ sample, respectively. Feature $s$ represents the \textit{scenario feature} that indicates which scenario the sample comes from based on some manual prior rules explicitly. Feature $c_i$ represents the $i_{th}$ feature in total $I$ \textit{common features} $\{c_1,...c_i,...c_I \}$. The goal of the CTR prediction~\cite{cheng2016wide,guo2017deepfm,varnali2021online} in the multi-scenario setting~\cite{wang2022causalint,sheng2021one,yang2022adasparse} is to learn a model $\hat{y}_j=f(x_j)$ with the provided training dataset $\mathcal{D}$.

\subsection{\name Overview}

\begin{figure*}[t]
    \centering
    \includegraphics[width=0.9\linewidth]{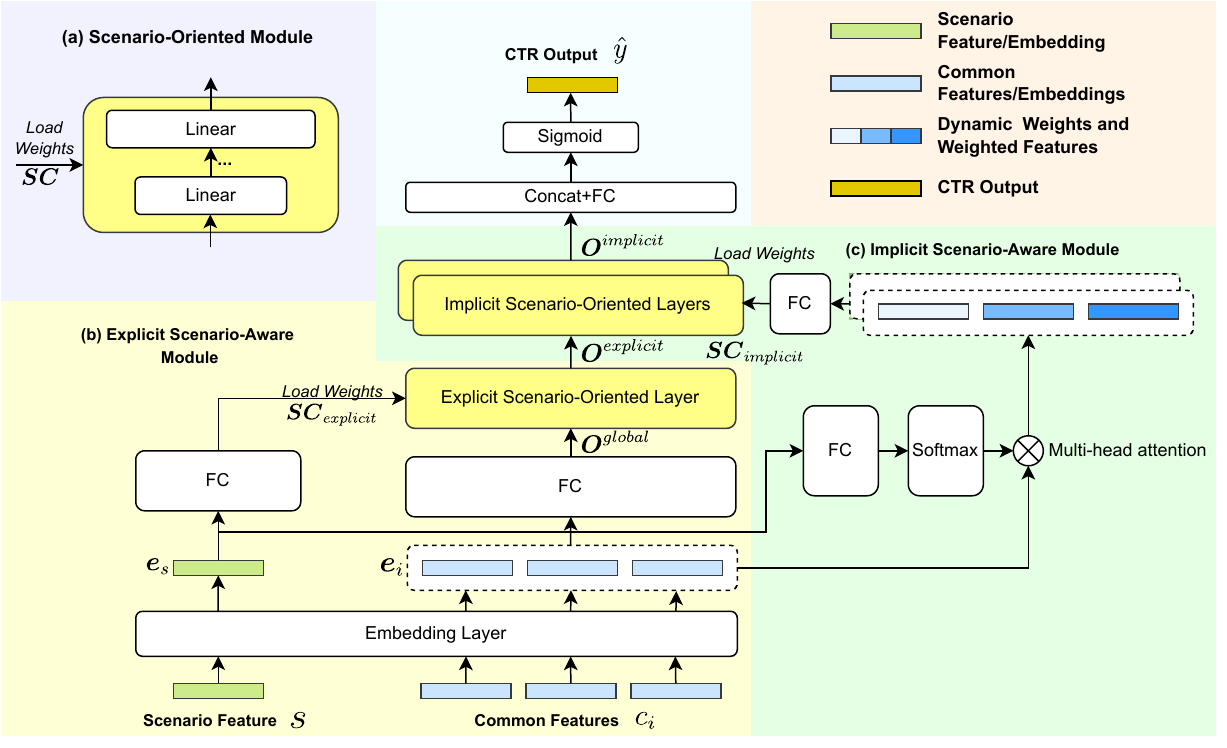}
    \caption{Overall structure of \name.}
    \label{fig:overview}
    \vspace{-5mm}
\end{figure*}

In this section, we present the overview architecture of \name with a hierarchical structure, illustrated in Figure~\ref{fig:overview}. An explicit scenario-oriented layer and several stacked implicit scenario-oriented layers are deployed to capture explicit and implicit information, respectively.
Specifically, HierRec first designs a basic \textbf{Scenario-Oriented Module} based on the dynamic weight to capture scenario-specific information, shown in Figure~\ref{fig:overview}~(a).
Then, an \textbf{Explicit Scenario-Aware Module} shown in Figure~\ref{fig:overview}~(b) is proposed to model coarse-grained explicit scenario information.
\name takes instance $\boldsymbol{x}_j$ as input and applies an embedding layer to transform sparse one-hot features, including both scenario feature and common features, into dense embeddings. 
The scenario feature embedding is fed into Fully Connected (FC) layers, and the output representation is used to parameterize the \textit{Explicit Scenario-Oriented Layer}, which is leveraged to model the explicit scenario.
Following an \textbf{Implicit Scenario-Aware Module} shown in Figure~\ref{fig:overview}~(c) is proposed to model fined-grained implicit scenario information.
A scenario-aware multi-head attention network is designed to perceive distinctive implicit patterns, which are further used to parameterize several \textit{Implicit Scenario-Oriented Layers} deployed parallelly for modeling complicated distribution from different perspectives.
Finally, the outputs of implicit scenario-oriented layers are concatenated and passed through the output layer for CTR prediction.

\subsection{Scenario-Oriented Module}
A key challenge in scenario modeling is how to provide a unified paradigm for modeling scenarios in different situations. To depict different explicit and implicit scenarios delicately in a parameter-efficient manner, inspired by the dynamic weight technique~\cite{apg}, HierRec proposes a scenario-oriented module based on the re-parameterization method to adaptively generate parameters depending on the given \textit{scenario condition}.
By instantiating the scenario-oriented module under different scenario conditions, HierRec can easily achieve explicit and implicit scenario-aware modeling while preserving scenario extensibility.
As illustrated in Figure~\ref{fig:overview}~(a), the scenario-oriented module is composed of several linear layers, in which the calculation of the $l_{th}$ linear layer could be expressed as:
\begin{equation}\label{equ:linear}
\boldsymbol{h}_{l+1}=\boldsymbol{W}_l \boldsymbol{h}_{l}+\boldsymbol{b}_l, \quad l \in[1, L],
\end{equation}
where $\boldsymbol{h}_{l}$ and $\boldsymbol{h}_{l+1}$ are the input and output, and $\boldsymbol{W}_l$ and $\boldsymbol{b}_{l}$ are its weights and bias, and $L$ is the number of the layers.

In order to model different scenarios according to different scenario-specific information, the network weights $\{\boldsymbol{W}_l\}_{l \in[1, L]}$ and $\{\boldsymbol{b}_{l}\}_{l \in[1, L]}$ are adaptively generated under different scenario conditions, which can be represented as:
\begin{equation}\label{equ:wb}
\boldsymbol{W}_l, \boldsymbol{b}_l= Reshape(\boldsymbol{SC})[l] \quad l \in[1, L],
\end{equation}
where $\boldsymbol{SC}$ is the given scenario conditions and $Reshape$ function splits $\boldsymbol{SC}$ into $L$ parts with the $l_{th}$ part for $\boldsymbol{W}_l$ and $\boldsymbol{b}_l$. In this paper, to simplify the design and reduce the number of parameters, referring to the bottleneck structure\cite{he2016deep,sandler2018mobilenetv2}, L is set to 2, where the first linear layer (i.e., bottleneck layer) contains fewer neurons, while the second linear layer contains more neurons.
Based on the basic scenario-oriented module, the following explicit scenario-oriented layer and implicit scenario-oriented layers are proposed to capture explicit and implicit information with different scenario conditions.

\subsection{Explicit Scenario-Aware Module}

To combine explicit and implicit scenario modeling, \name adopts a hierarchical structure to first model coarse-grained explicit scenario information and then conduct fine-grained implicit scenario modeling. Hence, a challenge in explicit scenario modeling is how to effectively and efficiently model multiple explicit scenarios. Therefore, the aim of this explicit scenario-aware module is to conduct explicit scenario-aware modeling based on pre-defined scenario identification. Specifically, the explicit scenario-aware module first embeds all features (including scenario and common features) into dense embeddings with shape $\mathbb{R}^{d}$ ($d$ is the embedding dimension) via an embedding layer:
\begin{equation}
\left\{\begin{aligned}
\boldsymbol{e}_i&=\boldsymbol{EM}_i \cdot Onehot(c_i), \quad i \in[1, I]\\
\boldsymbol{e}_s&=\boldsymbol{EM}_s \cdot Onehot(s),
\end{aligned}\right.
\end{equation}
where all the features are first transformed into one-hot vectors by $Onehot$ function and then transformed by the embedding matrices $\boldsymbol{EM}_i$ or $\boldsymbol{EM}_s$ according to the feature fields that they belong to. 

In order to save the model parameters and facilitate online inference, the common feature embeddings $\boldsymbol{E}_c=\{\boldsymbol{e}_1,...\boldsymbol{e}_i,...\boldsymbol{e}_I\}$ are concatenated and passed through a shared FC layer for dimension reduction and feature interaction modeling~\cite{wang2017deep}, obtaining global representation $\boldsymbol{O}^{global}$.
Afterward, the scenario embedding $\boldsymbol{e}_s$ is further dimensionally transformed through FC layers to yield explicit scenario condition $\boldsymbol{SC}_{explicit}$ for instantiating the explicit scenario-oriented layer:
\begin{equation}\label{equ:exg}
\boldsymbol{SC}_{explicit}= FC(\boldsymbol{e}_s).
\end{equation}
Here, FC layers contains $K$ layers, and the $k_{th}$ layer be:
\begin{equation}\label{equ:fc}
\boldsymbol{h}_{k+1}=\sigma (Dropout(BN( \boldsymbol{W}_k \boldsymbol{h}_{k}+\boldsymbol{b}_k))), \quad k \in[1, K],
\end{equation}
where $\sigma$ is the activation function~\cite{covington2016deep} of this layer, $Dropout$ is the dropout function~\cite{srivastava2014dropout}, $BN$ is the batch normalization function~\cite{ioffe2015batch} and $\boldsymbol{W}_k$ and $\boldsymbol{b}_k$ is the weight and bias of this layer. 

After instantiating the dynamic weights of the explicit scenario-oriented layer, the global dimension-reduced representation $\boldsymbol{O}^{global}$ is fed into the explicit scenario-oriented layer for explicit modeling. By doing this, we can obtain the representation under the current explicit scenario $\boldsymbol{O}^{explicit}$.

\subsection{Implicit Scenario-Aware Module}\label{sec:lhuc}

After the explicit scenario-aware modeling, \name intends to further excavate beneficial implicit patterns and realize a fine-grained implicit scenario-aware modeling under different explicit scenarios. Given the multitude of implicit patterns based on feature combinations and the fact that not all of them may necessarily be helpful for recommendations, it is important to identify beneficial implicit patterns adaptively and conduct scenario-aware modeling. This poses a significant challenge.
To fully perceive complex data distribution and identify important implicit patterns adaptively, \name proposes a scenario-aware multi-head attention structure.
Specifically, the explicit scenario embedding $\boldsymbol{e}_s$ is first fed into an FC layer, whose output representation is split and reshaped into multi-group weights, which are further normalized via the Softmax function~\cite{nelder1972generalized} to generate multi-group distributions.
This process can be expressed as:
\begin{equation}
\left\{\begin{aligned}
&\boldsymbol{weight}_{ori}=Reshape(FC(\boldsymbol{e}_s))\\
&\boldsymbol{weight}_{norm}[g]=Softmax(\boldsymbol{weight}_{ori}[g]), \\
&g \in[1, G]
\end{aligned}\right.
\end{equation}
where $\boldsymbol{weight}_{ori} \in \mathbb{R}^{G \times I}$, $\boldsymbol{weight}_{norm} \in \mathbb{R}^{G \times I}$ are the $G$ group weights before and after the Softmax normalization. $G$ is the number of attention heads and $I$ is the number of common features. 
By doing this, each weighted vector $\boldsymbol{weight}_{norm}[g]$ represents a kind of discovering implicit pattern over the common features, and each element in vector $\boldsymbol{weight}_{norm}[g]$ reflects the importance of the corresponding common feature under the current implicit scenario.
Finally, the weighted vectors $\boldsymbol{weight}_{norm}$ are multiplied with the common feature embeddings $\boldsymbol{E}_c$ in an element-wise manner, which can be denoted as:
\begin{equation}\label{equ:exg2}
\boldsymbol{IE} = \boldsymbol{weight}_{norm} \otimes \boldsymbol{E}_c,
\end{equation}
where $\boldsymbol{E}_c \in \mathbb{R}^{Id}$ is the concatenated common feature embeddings and $\boldsymbol{IE} \in \mathbb{R}^{G \times Id}$ is the $G$ groups identified implicit scenario representations.
By doing this, \name soft-selects several important implicit patterns adaptively, facilitating fine-grained modeling.

Afterward, implicit scenario representations $\boldsymbol{IE}$ are further dimensional transformed through a shared FC layer for obtaining $G$ disparate scenario conditions $\boldsymbol{SC}_{implicit}$, which can be deployed to instantiate $G$ implicit scenario-oriented layers. The $g_{th}$ scenario condition can be shown as:
\begin{equation}\label{equ:img}
\boldsymbol{SC}_{implicit}[g]= FC(\boldsymbol{IE}[g]), \quad g \in[1, G].
\end{equation}
Finally, the explicit representation $\boldsymbol{O}^{explicit}$ is then fed into the implicit scenario-oriented layers for implicit modeling, and the output representation for each implicit scenario-oriented layer can be denoted as $\boldsymbol{O}^{implicit}_g (g \in [1, G])$.

\subsection{Output Layer}

After the explicit and implicit scenario-aware modeling, the output representations of the $G$ implicit scenario-orient layers $\boldsymbol{O}^{implicit}_g (g \in [1, G])$ are concatenated together and passed through an FC layer with sigmoid function for CTR prediction $\hat{y}$, which be expressed as:
\begin{equation}\label{equ:outlayer}
\hat{y} = Sigmoid(FC(Concat(\boldsymbol{O}^{implicit}_1,...,\boldsymbol{O}^{implicit}_G))).
\end{equation}

The widely-used Binary Cross Entropy (BCE) loss~\cite{zhu2020ensembled,zhang2021deep} is deployed to measure the CTR accuracy with the prediction score $\hat{y}$ and the ground-truth label $y$, which is defined as follows:
\begin{equation}\label{equ:loss}
\mathcal{L}(\boldsymbol{\Phi})=-\frac{1}{|\mathcal{D}|} \sum_{j=1}^{|\mathcal{D}|}\left[y_j \log \hat{y}_j+\left(1-y_j\right) \log \left(1-\hat{y}_j\right)\right] ~.
\end{equation}

\section{Experiments}
In this section, we conduct experiments on two public datasets to investigate the following questions:
\begin{itemize}[leftmargin=*]
    \item \textbf{RQ1:} How does \name perform in comparison with multi-scenario recommendation baselines?
    \item \textbf{RQ2:} Is the designed hierarchical structure helpful in making predictions for different scenarios?
    \item \textbf{RQ3:} Is the inference efficiency of \name sufficient for online deployment requirements?
\end{itemize}


\subsection{Experimental Setup}
\subsubsection{Dataset}
We conduct experiments on two commonly-used datasets, i.e., Ali-CCP~\footnote{https://tianchi.aliyun.com/dataset/408}~\cite{ma2018entire} and KuaiRand~\footnote{https://kuairand.com/}~\cite{gao2022kuairand}. For Ali-CCP which has a training set and a test set, following~\cite{xu2022recommendation} we split the training set into training/validation sets with an 8:2 ratio. For Ali-CCP, the classification of explicit scenarios follows the settings of the official instruction and previous work~\cite{wang2022causalint}, which is expressed by the discrete feature ``\textit{301}'' indicating a categorical expression of recommendation position. For KuaiRand, to facilitate evaluation, we select the top-5 pre-defined scenarios with the most data for evaluation and split the dataset into training/validation/test sets with an 8:1:1 proportion~\cite{zhang2022leaving}.
We follow the settings of the official description~\cite{gao2022kuairand} to divide explicit scenarios with discrete feature ``\textit{tab}'', which indicates the interaction scenario such as the recommendation page or main page of the Kuaishou App\footnote{https://www.kuaishou.com/cn}.
The statistics of the two datasets are summarized in Table~\ref{tab:statistics}. 

\begin{table}[t]
\setlength\tabcolsep{4.0pt}
  \centering
  \caption{Statistics of evaluation datasets}
    \begin{tabular}{cccccc}
    \toprule
    \multirow{2}{*}{Dataset} & \multirow{2}{*}{\#Scenarios} & \multicolumn{1}{c}{\multirow{2}{*}{\#Features}} & \multicolumn{3}{c}{Instances(M)} \\
\cmidrule{4-6}          &       &       & \multicolumn{1}{c}{Train} & \multicolumn{1}{c}{Val} & \multicolumn{1}{c}{Test} \\
    \midrule
    Ali-CCP & 3     & 23    & 38.07 & 4.23  & 43.02 \\
    KuaiRand & 5     & 37    & 5.28  & 0.66  & 0.66 \\
    \bottomrule
    \end{tabular}%
  \label{tab:statistics}%
  \vspace{-3mm}
\end{table}%

\subsubsection{Baseline}
To verify the effectiveness of the proposed approach, we compare \name with the following baselines: 
\begin{itemize}[leftmargin=*]
    \item \textbf{Shared Bottom} shares the embedding layer and bottom FC layers, and several scenario-specific FC layers are adopted for each scenario.
    \item \textbf{MMoE}~\cite{ma2018entire} implicitly models task relationships for multi-task learning. Here we treat different scenarios as different tasks and apply scenario-specific towers and gating networks for each scenario.
    \item \textbf{PLE}~\cite{tang2020progressive} uses a progressive layered extraction for multi-task learning. Similar to MMoE, we apply scenario-specific experts and towers for each scenario.
    \item \textbf{STAR}~\cite{sheng2021one} utilizes scenario-specific tower networks to learn scenario-specific information, and a shared network to learn shared information.
    \item \textbf{AdaSparse}~\cite{yang2022adasparse} utilizes scenario embeddings as a unique input to implement scenario-aware neuron-level weighting and then adaptively learns different sparse structures for each scenario.
\end{itemize}


\subsubsection{Implementation Details}

The widely used metrics of AUC and Logloss are deployed for evaluation. Specifically, a higher AUC value or a lower Logloss at the ``0.001'' level indicates significantly better performance~\cite{guo2017deepfm}. Besides, RelaImpr~\cite{shen2021sar,yan2014coupled} is also applied to measure the relative improvement between \name and best baselines:

\begin{equation}
\text { RelaImpr }=\left(\frac{A U C(\text {\name})-0.5}{A U C(\text {Best baseline})-0.5}-1\right) \times 100 \%.
\end{equation}

For a fair comparison, we fix the embedding size of each feature at 16, the batch size at 2000, and the optimizer is the commonly used ``Adam Optimizer"~\cite{kingma2014adam}. Simple grid searches are performed for all the adjustable hyper-parameters of \name and baselines. For FC layers, the number of layers is searched from 1 to 5, and neurons at each layer from \{16, 32, 64, 128\}. Besides, we run each experiment 10 times with the optimal parameters searched and report the average performance. For ease of reproduction, we provide the source code for the experiments conducted using the Ali-CCP and KuaiRand datasets in the supplementary materials. Additionally, we have provided data samples from both datasets for reference purposes.

\subsection{Overall Performance (RQ1)}

This subsection gives an overall comparison between \name and different baselines, whose results are depicted in Table~\ref{tab:baseline}. From this we can conclude that: 


\begin{itemize}[leftmargin=*]
    \item Multi-task based models (Shared Bottom, MMoE, PLE) achieve acceptable results on both datasets, which demonstrates that benefiting from the task sharing and exclusive mechanisms, multi-task learning based methods can also be applied to multi-scenario recommendations.
    MMoE outperforms Shared Bottom due to the modeling of task relations and better sharing design with gating networks.
    Besides, PLE outperforms the other two models, illustrating the effectiveness of refined information isolation in scenario-shared and scenario-specific modules and the progressive routing mechanism for information extraction.

    \item  Multi-scenario based models (STAR, AdaSparse) achieve better performance than multi-task based models, elaborating the significance of effectively modeling the differences and associations within different explicit scenarios. 
    In addition, from the overall performance, AdaSparse outperforms STAR due to the fine-grained scenario modeling at the neuron level, which contributes to precisely recognizing the scenario distinctions.

    
    \item  \name outperforms all the baselines in both scenario-individual and overall performance by a significant margin, showing superior prediction capabilities and proving the effectiveness of combining explicit and implicit scenario modeling. The multi-head implicit modeling design contributes to perceiving complicated distributions and achieving fine-grained modeling. Additionally, the improvements of \name in KuaiRand are less than that in Ali-CCP. We attribute this distinction to the complexity of different data distributions.
    For complex scenarios, the hierarchical modeling of \name brings superior modeling ability and uncovers more intricate correlations within scenarios, thus achieving remarkable improvements.
    
\end{itemize}

\begin{table*}[t]
  \setlength\tabcolsep{0.5pt}
  \centering
  \caption{Performance comparison of \name and baselines, where sce\_$d$ indicates the evaluation in the $d$-th scenario. Boldface denotes the highest score and underline indicates the best result of all baselines. ``\textbf{{\Large *}}'' indicates the statistically significant improvements (i.e., two-sided t-test with $p<0.05$) over the best baseline. $\uparrow$: higher is better; $\downarrow$: lower is better.}
    \begin{tabular}{ccccccccc||cccc}
    \toprule
    \multirow{3}{*}{Approach} & \multicolumn{8}{c||}{Performance for Each Scenario (AUC $\uparrow$)}                            & \multicolumn{4}{c}{Overall Performance} \\
\cmidrule{2-13}          & \multicolumn{3}{c}{Ali-CCP} & \multicolumn{5}{c||}{KuaiRand}        & \multicolumn{2}{c}{Ali-CCP} & \multicolumn{2}{c}{KuaiRand} \\
\cmidrule{2-13}          & sce\_1 & sce\_2 & sce\_3 & sce\_1 & sce\_2 & sce\_3 & sce\_4 & sce\_5 & AUC  $\uparrow$ & Logloss $\downarrow$& AUC $\uparrow$  & Logloss $\downarrow$\\
    \midrule
    Shared Bottom & 0.6094  & 0.5545  & 0.6064  & 0.7298  & 0.7183  & 0.7187  & 0.7904  & 0.7565  & 0.6030  & 0.2062  & 0.7757  & 0.5453  \\
    MMoE  & 0.6181  & 0.5727  & 0.6123  & 0.7292  & 0.7199  & 0.7153  & 0.7794  & 0.7553  & 0.6107  & 0.1635  & 0.7776  & 0.5444  \\
    PLE   & 0.6154  & 0.5919  & 0.6126  & 0.7285  & 0.7221  & 0.7188  & 0.7902  & 0.7661  & 0.6133  & 0.1621  & 0.7784  & 0.5427  \\
    STAR  & \underline{0.6187}  & 0.5954  & 0.6132  & \underline{0.7323}  & 0.7205  & \underline{0.7204}  & 0.7903  & 0.7772  & 0.6149  & 0.1622  & 0.7802  & 0.5415  \\
    AdaSparse & 0.6186  & \underline{0.5970}  & \underline{0.6164}  & 0.7320  & \underline{0.7301}  & 0.7197  & \underline{0.7971}  & \underline{0.8184}  & \underline{0.6165}  & \underline{0.1620}  & \underline{0.7815}  & \underline{0.5384}  \\
    HierRec & \textbf{0.6253* } & \textbf{0.6046* } & \textbf{0.6228* } & \textbf{0.7351* } & \textbf{0.7324* } & \textbf{0.7250* } & \textbf{0.8005* } & \textbf{0.8442* } & \textbf{0.6237* } & \textbf{0.1614* } & \textbf{0.7847* } & \textbf{0.5376* } \\
    \midrule
    RelaImpr & 5.56\%  & 7.84\%  & 5.50\%  & 1.21\%  & 1.00\%  & 2.09\%  & 1.14\%  & 8.10\%  & 6.18\%  & -  & 1.14\%  & -  \\
    \bottomrule
    \end{tabular}%
  \label{tab:baseline}%
\end{table*}%

\begin{figure}[h]
	\centering
        \subfigure[AUC.]{
		\label{fig:ab1}
		\includegraphics[width=0.47\linewidth]{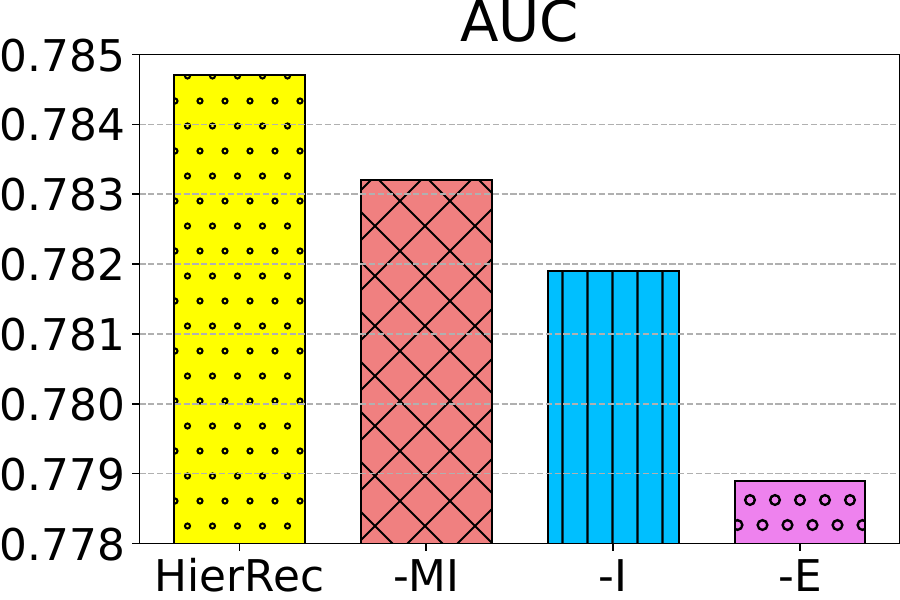}}
        \subfigure[Logloss.]{
		\label{fig:ab2}
		\includegraphics[width=0.47\linewidth]{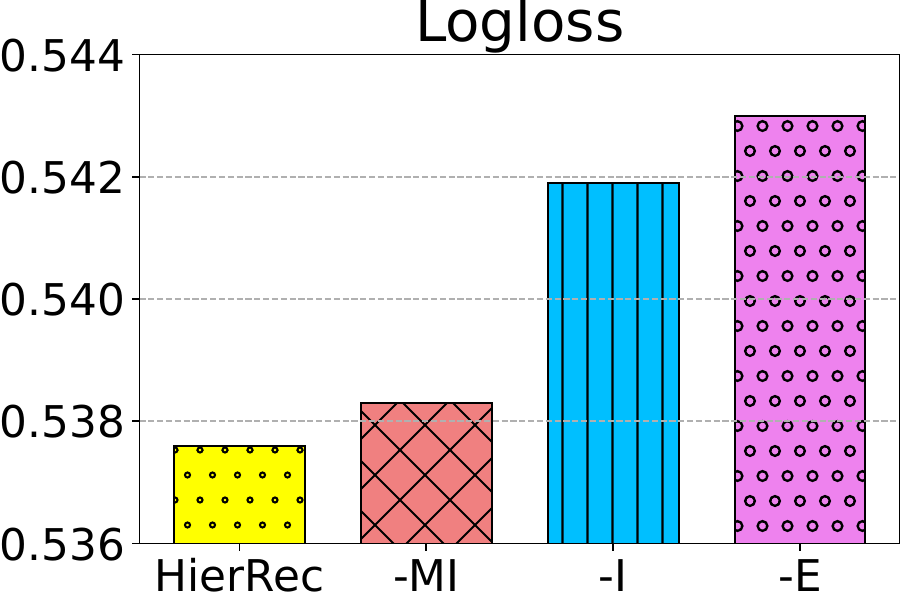}}
\caption{Ablation Study of overall scenarios performance on the KuaiRand dataset.}
\label{fig:ablation}
\vspace{-4mm}
\end{figure}

\subsection{Ablation Study (RQ2)}
This section presents the ablation study of our proposed HierRec model. Specifically, we compare HierRec with the following alternatives on the KuaiRand dataset:

\begin{itemize}[leftmargin=*]
    \item \textbf{w/o multi-head attention (-MI)}: with only one head in the implicit scenario-aware module;
    \item \textbf{w/o implicit layers (-I)}: without implicit scenario-oriented layers for implicit modeling; 
    \item \textbf{w/o explicit layers (-E)}: without explicit scenario-oriented layers for explicit modeling.
\end{itemize}

Based on the results in Figure~\ref{fig:ablation}, we could conclude that both the explicit and implicit scenario modeling play an important role for HierRec. Besides, explicit modeling is more prominent, which is also the selection motivation of existing work~\cite{sheng2021one,yang2022adasparse}. This is because samples in different explicit scenarios often exhibit significant distribution differences, arising from their unique positions or presentation methods (e.g., advertising slot). Modeling explicit scenarios adequately allows for capturing these explicit scenario-specific differences~\cite{sheng2021one}. However, the improvement brought by implicit modeling is non-negligible as it could further uncover more intricate correlations among samples through the exploration of feature-based implicit patterns.
Moreover, the multi-head implicit modeling can perceive complicated data distribution sufficiently, thus conducive to fine-grained implicit modeling.




\begin{table}[h]
  \centering
  \caption{Inference time of \name and baselines on the whole test sets. The calculation of the increase percentage is based on the baseline which takes the most time.}
    \begin{tabular}{ccc}
    \toprule
    \multirow{2}[4]{*}{Approach} & \multicolumn{2}{c}{Inference Time (Seconds)} \\
\cmidrule{2-3}          & Ali-CCP (43M) & KuaiRand (1.5M) \\
    \midrule
     Shared  Bottom & 518.86  & 9.24  \\
     MMoE & 506.32  & 9.19  \\
     PLE  & 558.88  & 10.22  \\
     STAR & 503.56   & 8.81  \\
     AdaSparse & 510.37   & 8.90  \\
    HierRec (Ours) & 572.44  & 10.29  \\
    \midrule
    Increase & 2.43\% & 0.68\% \\
    \bottomrule
    \end{tabular}%
  \label{tab:infer}%
  \vspace{-5mm}
\end{table}%

\subsection{Inference Efficiency Analysis (RQ3)}
In practical applications, the inference efficiency of CTR models is a significantly important index due to the essential need for real-time response in recommender systems. Therefore, to answer RQ3, this subsection presents a comparison of inference time between \name and other baselines on the test set of Ali-CCP and KuaiRand. The experiments are conducted on NVIDIA GeForce RTX 3060 GPU over the entire test set (43 million testing samples for Ali-CCP and 1.5 million testing samples for KuaiRand), whose results are summarized in Table~\ref{tab:infer}. 
Based on the results, it can be concluded that \name's inference time increases slightly compared to other baselines due to the detailed multi-head implicit scenario-aware modeling. The increase in inference time is minor and acceptable for industrial applications.


\section{Application: Online Advertising Platform}


\subsection{Scenario Description \& Experimental Setting}
In this section, we deploy \name in the Lead Ads Recommendation in a mainstream online advertising platform to verify its effectiveness. 
Lead Ads Recommendation Platform contains several major industries, such as Automobile, Finance, and Real Estate, where industry identification is used as the explicit scenario feature to divide scenarios explicitly. An example of the Automobile industry in Lead Ads Recommendation is presented in Figure~\ref{fig:form_example}.
Besides, more than 80 common features are used to divide scenarios implicitly, including user profiles (e.g., gender), ads features (e.g., category), as well as contextual features (e.g., ad slot).

For the categorical features, the feature embeddings are learned via embedding look-up, while the numerical feature embeddings are generated via the AutoDis~\cite{autodis}.
We collect and sample one month of user behavior record to train baseline models, including single-scenario models (FiBiNet~\cite{fibinet}, DCN~\cite{wang2017deep}) and multi-scenario/task models (DFFM~\cite{dffm}, MMoE~\cite{ma2018entire}, PLE~\cite{tang2020progressive}) which are widely used in industrial recommender systems.

\subsection{Experimental Results}
\subsubsection{Offline and Online Results}
\begin{figure}
    \centering
    \subfigure[Automobile industry Ads]{
    \centering
    \label{fig:form_example}
    \includegraphics[width=0.35\linewidth]{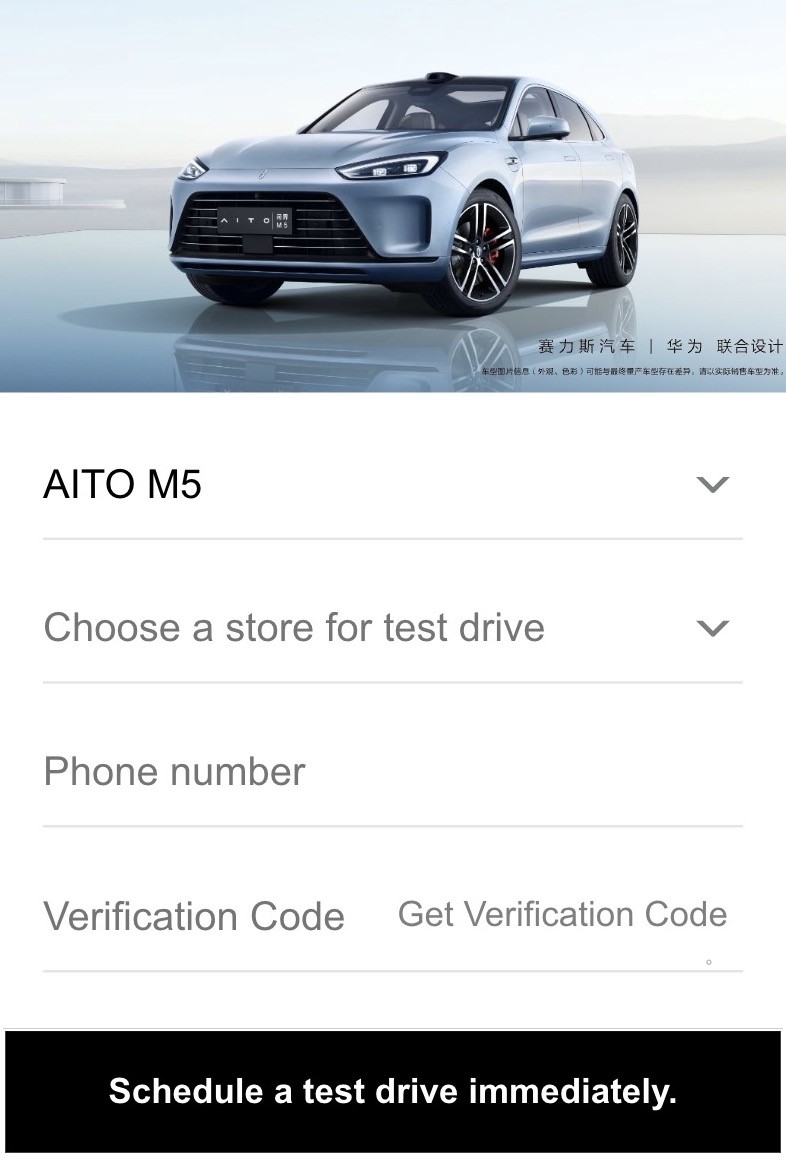}
    }
    \subfigure[Performance comparison]{
    \centering
    \label{fig:industrial_res}
    \includegraphics[width=0.59\linewidth]{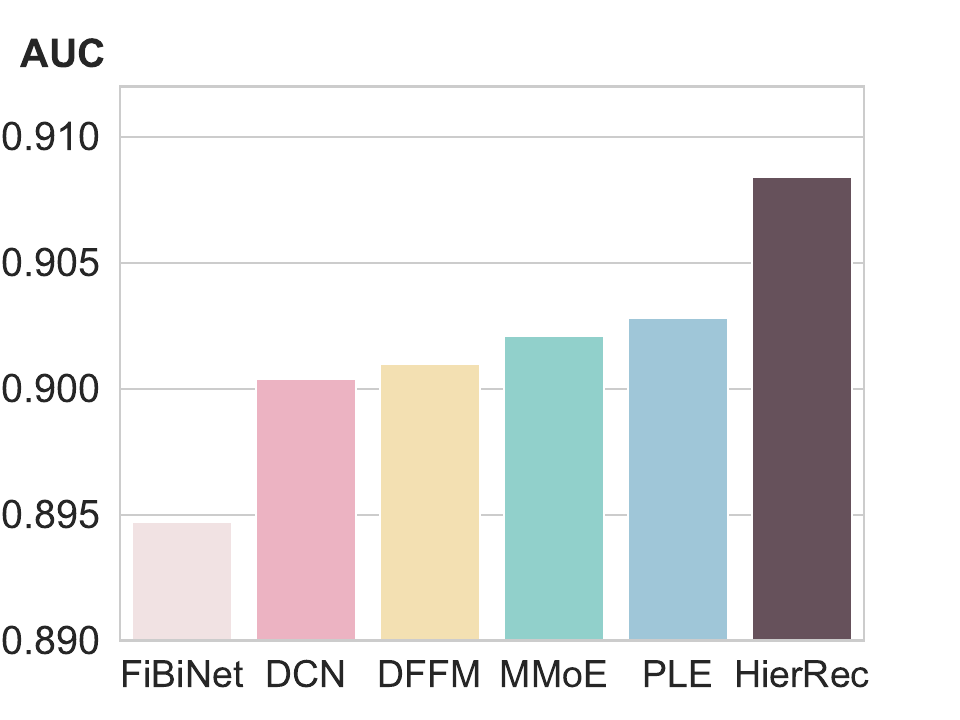}
    }

\caption{An example of the \textbf{Automobile}
industry in Lead Ads and the performance comparison over baselines.}

\end{figure}

The offline performance comparison on the large-scale industry dataset is presented in Figure~\ref{fig:industrial_res}. We can observe that our proposed \name outperforms all the baselines including single-scenario and multi-scenario/task models by a significant margin, verifying its effectiveness.

To verify the performance of HierRec online, we conduct a two-week online A/B testing on the Lead Ads Recommendation, whose results are shown in Table \ref{tab:online_res_tab}. Compared with the optimal baseline, which is a highly-optimized deep multi-scenario model, eCPM (effective cost per mile) is improved by 10.33\% and the predicted bias is reduced by 6.81\%. As a platform to recommend ads for the users, the higher eCPM means better online advertising effectiveness and the lower bias implies more accurate prediction and cost control, which is critical for the advertisers. 
Besides, \name has comparable inference efficiency with other models as shown in Table~\ref{tab:infer}, which demonstrates that \name is suitable for industrial applications.

\begin{table}[t]
  \centering
  \caption{The online A/B testing results of HierRec compared with the optimal baseline.}
  \vspace{-3mm}
    \begin{tabular}{ccc}
    \toprule
     & eCPM  & predicted bias  \\
    \midrule
    Improvements & +10.33\%  & -6.81\%  \\
    \bottomrule
    \end{tabular}%
  \label{tab:online_res_tab}%
\end{table}%

\subsubsection{Implicit Scenario Analysis}
\begin{figure}[t]
	\centering
\includegraphics[width=0.8\linewidth]{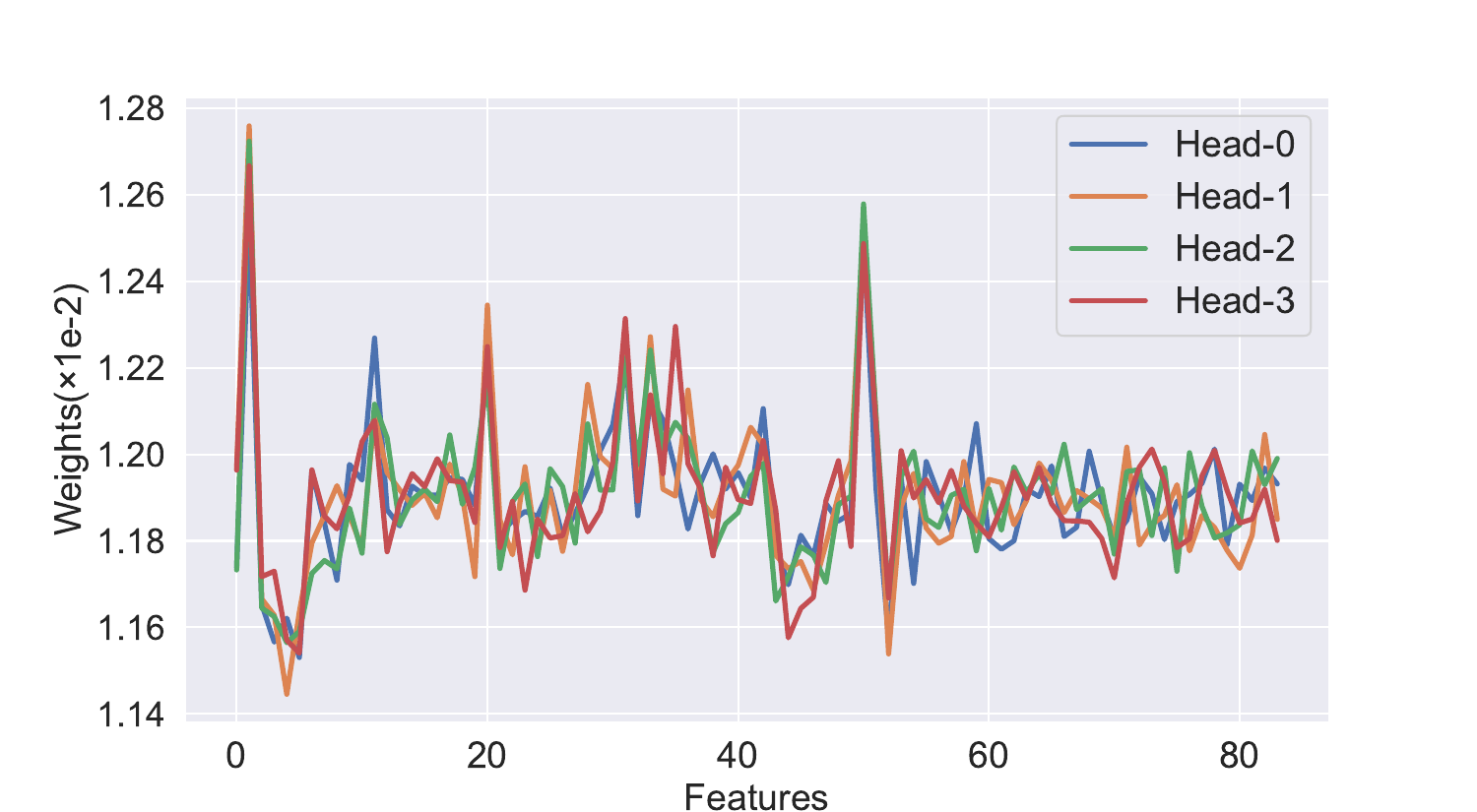}
\caption{Feature weights of the industrial dataset. Each count on the horizontal axis represents a common feature. Head-$g$ represents the weights of the $g_{th}$ attention head in implicit scenario-aware modeling.}
\label{fig:weight}
\end{figure}

We visualize the weights $\boldsymbol{weight}_{norm}$ of the multi-head implicit scenario-aware modeling on the industrial dataset, as illustrated in Figure~\ref{fig:weight}. From the results in Figure~\ref{fig:weight}, it is evident that different attention heads assign varying weights to common features, enabling the perception and discovery of beneficial patterns in complex data distribution. In addition, several features consistently receive higher weights than others, underscoring the importance of these features for implicit scenario modeling and decision-making, thus compensating for the inadequacy of explicit modeling.


\section{Related Work}
This section offers a brief overview of multi-scenario recommendations~\cite{jiang2022adaptive,zang2022survey} (a.k.a., multi-domain recommendations).
Currently, existing multi-scenario models can be divided into two categories: Tower-based models and Dynamic Weight (DW) models, whose abstract structures are depicted in Figure~\ref{fig:compare}. Specifically, tower-based models utilize a common network to represent scenario-shared information, upon which multiple sub-towers are built to capture scenario-specific details. 
Multi-task models belong to this category, such as Shared Bottom, MMoE~\cite{ma2018entire}, and PLE~\cite{tang2020progressive}, which design task-sharing and task-specific networks to model task relations.
Besides, STAR~\cite{sheng2021one} utilizes several independent towers to learn scenario-specific information, and a shared network to learn global information. It also leverages element-wise multiplication to establish connections between the tower and shared networks.
CausalInt~\cite{wang2022causalint} further eliminates negative transfers among different tower networks. With the design of a causal intervention method, CausalInt is able to selectively utilize the information from different scenarios to construct scenario-aware estimators in a unified model.


However, the complete isolation design between towers hinders the modeling of scenario correlations and also suffers from poor generalization and compatibility. To overcome these limitations, DW-based methods have been proposed by generating dynamic parameters adaptively for each scenario in a parameter-efficient manner~\cite{zhang2022leaving, yang2022adasparse}. M2M~\cite{zhang2022leaving} proposes a meta-unit, which uses scenario information to generate dynamic weights as parameters of different networks to realize multi-scenario and multi-task learning simultaneously. AdaSparse~\cite{yang2022adasparse} utilizes scenario embeddings as unique input to implement scenario-aware neuron-level weighting so that it can adaptively learn sparse structures.
However, all these existing multi-scenario models only consider coarse-grained explicit scenario modeling that depends on pre-defined scenario identification based on some manual prior rules, which is biased and sub-optimal. Therefore, to realize fine-grained modeling over complex data distribution, \name is purposed with a hierarchical structure to model explicit and implicit scenarios jointly.



\section{Conclusion}
In this paper, we propose a scenario-aware hierarchical dynamic network \name to conduct explicit and implicit scenario modeling simultaneously. 
Specifically, a basic scenario-oriented module is designed to capture scenario-specific information. Then the stacked explicit and implicit scenario-aware modules are proposed to model explicit and implicit scenario information in a hierarchical manner. Moreover, the multi-head implicit modeling design can perceive distinctive patterns effectively and achieve fine-grained modeling.
Experiments on two public datasets and applications on a mainstream online advertising platform demonstrate the effectiveness of the proposed \name.


\bibliography{aaai24}

\end{document}